\documentclass[1p,twoside]{article}
\usepackage[english]{babel}
\usepackage{amsmath,amsfonts,amssymb,amsthm}
\usepackage{graphicx}
\usepackage{xcolor}
\usepackage{float}
\usepackage[numbers]{natbib}
\usepackage{xr-hyper}
\usepackage[hidelinks]{hyperref}
\usepackage{soul}
\usepackage{enumerate}
\usepackage{enumitem}
\usepackage{multirow}
\usepackage{bm}
\usepackage[leftcaption]{sidecap} 
\usepackage{booktabs,makecell,tabularx}
\usepackage{lipsum}
\allowdisplaybreaks
\usepackage[numbers]{natbib}
\allowdisplaybreaks
\usepackage{comment}
\usepackage{lscape}
\usepackage{anyfontsize}
\usepackage{adjustbox}
\usepackage{afterpage}

\usepackage[left=2cm,right=2cm,top=2cm,bottom=2cm]{geometry}

\title{Comparing Lasso and Adaptive Lasso in
High-Dimensional Data: A Genetic Survival
Analysis in Triple-Negative Breast Cancer}

\date{}
\author{Pilar Gonz\'alez-Barquero \footnote{uc3m-Santander Big Data Institute, University Carlos III of Madrid. E-mail address: \url{mariapilar.gonzalez@uc3m.es}. ORCID: 0009-0009-3905-8094.} 
\and
Rosa E. Lillo \footnote{uc3m-Santander Big Data Institute, University Carlos III of Madrid. E-mail address: \url{lillo@est-econ.uc3m.es}. ORCID: 0000-0003-0802-4691.}
\and \'Alvaro M\'endez-Civieta \footnote{Department of Biostatistics, Columbia University, New York. E-mail address: \url{am5490@cumc.columbia.edu}. ORCID: 0000-0003-2044-4170.}}
\begin{document}
\maketitle

\begin{abstract}
In high-dimensional survival analysis, effective variable selection is crucial for both model interpretation and predictive performance. This paper investigates Cox regression with lasso and adaptive lasso penalties in genomic datasets where covariates far outnumber observations. We propose and evaluate four weight calculation strategies for adaptive lasso specifically designed for high-dimensional settings: ridge regression, principal component analysis (PCA), uni\-variate Cox regression, and random survival forest (RSF) based weights. To address the inherent variability in high dimensional model selection, we develop a robust procedure that evaluates performance across multiple data partitions and selects variables based on a novel importance index. Extensive simulation studies demonstrate that adaptive lasso with ridge and PCA weights significantly outperforms standard lasso in variable selection accuracy while maintaining similar or better predictive performance across various correlation structures, censoring proportions (0-80\%), and dimensionality settings. These improvements are particularly pronounced in highly-censored scenarios, making our approach valuable for real-world genetic studies with limited observed events. We apply our methodology to triple-negative breast cancer data with 234 patients, over 19500 variables and 82\% censoring, identifying key genetic and clinical prognostic factors. Our findings demonstrate that adaptive lasso with appropriate weight calculation provides more stable and interpretable models for high-dimensional survival analysis.
\end{abstract}


\section{Introduction}\label{intro}

The application of genetic information in statistics has seen significant growth due to advancements in genomic technologies and the increasing availability of large-scale genetic data. This has led to the development of numerous methodologies specifically designed to analyze these complex, high-dimensional datasets. Genetic data analysis has various medical applications, including identifying genetic associations for disease risk prediction and assessing treatment efficacy.  Understanding the genetic foundations of diseases facilitates the development of more targeted therapies, as knowledge of an individual's genetic profile can determine the selection of more effective tailored treatments . Lee \textit{et al.} \cite{lee2017} demonstrated this potential in their work on variable selection for high-dimensional genomic data, showing how appropriately selected genetic signatures can dramatically improve survival predictions compared to conventional clinical factors alone. In addition, other examples of these applications are developed in Hastie $\&$ Tibshirani \cite{Hastie} and Chun $\&$ Keleş \cite{Chun} among others.

Cancer represents a prime example where genetic information plays an essential role in both understanding disease mechanisms and developing personalized treatments. Breast cancer is the most prevalent cancer among women, constituting $25\%$ of all cases (Ferlay \textit{et al.}, \cite{ferlay}). Within breast cancer, $15$-$20\%$ of cases are classified as triple-negative breast cancer (TNBC), characterized by the absence or low concentration of the receptors ER, PR, HER-2/neu (Dent \textit{et al.}, \cite{cancer}). TNBC is particularly aggressive, showing rapid growth and higher recurrence rates post-treatment, compared to other breast cancer subtypes, resulting in significantly lower survival rates. Given these challenges, our research focuses on identifying clinical and genetic features that influence TNBC patient survival using robust statistical approaches.

Survival analysis is the branch of statistics designed to study time-to-event data, with nume\-rous applications particularly in medicine. It plays a vital role in understanding and predicting patient survival under varying conditions, aiding medical professionals in making decisions and developing treatment strategies. Our work focuses on the Cox proportional hazard model or Cox regression (Cox, \cite{Cox_1972}), a widely employed method in survival analysis. An important challenge emerges when analyzing high-dimensional datasets like the TNBC data described above. Although having access to these large datasets provides a large amount of information, handling such large volumes of data adds significant complexity and difficulty to the decision-making process. Additionally, survival data often includes a large percentage of censored observations, where exact survival times remain unknown and only lower bounds are available. To contextualize the case study that motivates this research, the dataset provided by the Gregorio Marañón General University Hospital (GMGUH) is a very complex high-dimensional dataset with 82\% of censored observations.

In high-dimensional survival contexts, standard Cox regression models become unfeasible. Regularization or variable selection methods are essential to address this limitation, enabling model resolution and enhancing both interpretability and predictive accuracy. The Least Absolute Shrinkage and Selection Operator (lasso) represents one of the most influential regularization methods. It was introduced by Tibshirani \cite{Tibshirani_1996} and later adapted to Cox regression by Tibshirani \cite{Tibshirani_1997}. The lasso employs an $L_1$-norm that shrinks some of the coefficients to exactly zero, effectively performing variable selection.  This has spawned several extensions, including the group lasso penalization (Yuan $\&$ Li, \cite{Yuan}), hierarchical lasso (Zhou $\&$ Zhu, \cite{Zhou}) and sparse group lasso (Friedman \textit{et al.} \cite{Friedman}) among others, all of which were subsequently extended to Cox regression by Kim \textit{et al.} \cite{Kim}, Wang \textit{et al.} \cite{Wang} and Simon \textit{et al.} \cite{Simon} respectively. 

Despite its popularity, lasso has theoretical and practical limitations. When applied to genomic data, Mendez-Civieta \textit{et al.} \cite{Mendez} showed that lasso results were not stable, producing inconsistent variable selection across different data partitions. This instability poses serious challenges for researchers seeking reliable gene identification in clinical studies. From a theoretical perspective, Zou \cite{Zou} proved that lasso fails to satisfy the oracle property due to its inherent estimation bias. The constant penalization rate in lasso shrinks all coefficients uniformly, regardless of their true importance, leading to biased estimates for large coefficients and suboptimal variable selection.

Our research focuses on adaptive lasso, an alternative to standard lasso proposed by Zou \cite{Zou} that addresses these limitations. The key innovation of adaptive lasso is its assignment of different weights to each variable in the penalty term, increasing model flexibility. Importantly, Zou \cite{Zou} proved that adaptive lasso satisfies the oracle property defined by Fan $ \&$ Li \cite{Fan} when weights are based on $\sqrt{n}$-consistent estimators. This means that under certain conditions, adaptive lasso can correctly identify the true underlying model. The extension of this penalization to Cox regression was later developed by Zhang $\&$ Lu \cite{Zhang}. 

However, applying adaptive lasso to high-dimensional data presents  challenges, particularly in weight calculation. Most established weight calculation procedures for adaptive lasso, such as those introduced by Zou \cite{Zou} using ordinary least squares, become unfeasible in high-dimensional scenarios. This paper makes several significant contributions in addressing these challenges. First, we propose novel weight calculation strategies for adaptive lasso specifically designed for high-dimensional survival contexts. These include methods based on principal component analysis (Jolliffe $\&$ Cadima, \cite{Jollife}), ridge regression (Hoerl $\&$ Kennard, \cite{Hoerl_1970}), univariate Cox regressions, and Random Survival Forest (RSF) (Ishwaran \textit{et al.}, \cite{Ish}). Second, we introduce a method to simulate survival data with predefined censoring proportions, enabling realistic and controlled evaluation environments —crucial for testing models intended for highly censored genetic survival data. Third, we present an innovative approach for selecting the best predictive model by leveraging multiple partitions of the training data, and formulate a model selection procedure that reduces variability in variable selection in high-dimensional settings. Fourth, we conduct an extensive simulation study that incorporates various realistic covariate configurations (correlation structures, censoring percentages, and proportions of important variables) to comprehensively assess the proposed methods. Finally, we apply these methods to the motivating TNBC data, identifying key clinical and genetic factors influencing patient survival—findings with potential impact on treatment strategies.

The rest of this article is structured as follows. In Section \ref{pencox}, the theoretical concepts for Cox regression with lasso and adaptive lasso penalization techniques are introduced. In Section \ref{weights}, we introduce our four novel weight calculation strategies specifically designed for adaptive lasso in high-dimensional survival settings. Section \ref{cind} examines appropriate performance measures for Cox regression models, focusing on concordance-based metrics that accommodate right-censored data. Section \ref{best_model} presents our proposed model selection methodology that addresses variability in high-dimensional settings through an importance index —a significant contribution of this work. Section \ref{simul} details the extensive set of simulation studies where lasso and adaptive lasso survival models are fitted and subsequently evaluated.. In Section \ref{cstudy}, we apply our methodology to the triple-negative breast cancer dataset. Finally, the results obtained are discussed in Section \ref{conclusion}, and directions for future research in high-dimensional survival analysis are provided.

\section{Penalized Cox regression}\label{pencox}

In survival analysis, the main variable of interest is the survival time (T) which represents the time until an event occurs and can be characterized by the hazard function:
\begin{equation*}
h(t)=\lim_{dt\to 0} \frac{P(t\leq T < t+dt | T \geq t)}{dt}.
\end{equation*}
This function represents the probability of the event occurring during any given time point $t$. A common challenge in survival analysis is the presence of right-censored data, when only a lower bound for the survival time of some individuals is known. Whether the data is right-censored or not is denoted by an indicator $\delta_i$ for each individual $i$ that takes the value $0$ for censored data and $1$ for uncensored data.

The objective in survival analysis is to investigate the relationship between survival time and several observed covariates, identifying key risk factors that may affect the survival time. In the context of censored data, the Cox regression model is a fundamental approach widely adopted in medical research and clinical trials. The value of the hazard function conditioned on the covariates is given by
\begin{equation}\label{like}
h(t|X)=h_0(t)\exp\left(\bm{\beta}^T \bm{X}\right),
\end{equation}
where $\bm{X}^T=(X^1,\ldots,X^p)$ represents the vector of $p$ covariates, $h_0(t)$ is the baseline hazard function when all covariates are zero and $\bm{\beta}=(\beta_1,\ldots,\beta_p)^T$ is the column vector of regression coefficients, which is unknown and needs to be estimated. Applying a logarithm transformation to Equation \eqref{like}, this regression model can be transformed into a linear form 
\[\log\frac{h(t|X)}{h_0(t)}=\bm{\beta}^T \bm{X},\]
facilitating interpretation of the model parameters - each coefficient $\beta_i$ represents the log hazard ratio associated with a one-unit increase in the corresponding covariate $X^i$. The partial likelihood function for Cox regression introduced by Cox \cite{Cox_1975} is utilized for estimating regression coefficients. It is defined as
\begin{equation}\label{partiallikelihood}
L(\bm{\beta})=\prod_{j=1}^n\frac{\exp\left(\bm{\beta}^T \bm{X}_j\right)}{\sum_{k\in R_j}\exp\left(\bm{\beta}^T \bm{X}_k\right)},
\end{equation}
where $n$ is the number of observed individuals, $\bm{X}_j$ is the vector of covariates associated to individual $j$ and $R_j$ represents the individuals that have survived or have not been censored by time $t_j$, which is the observed time for the $j$-th individual. Applying the logarithm to Equation \eqref{partiallikelihood}, the log partial likelihood function is 
\begin{equation}\label{loglik}
l(\bm{\beta})=\sum_{j=1}^n\left(\bm{\beta}^T \bm{X}_j\right)-\sum_{j=1}^n\log\left(\sum_{k\in R_j}\exp(\bm{\beta}^T \bm{X}_k)\right),
\end{equation}
and the maximum partial likelihood estimator of $\beta$ is given by
\begin{equation*}
   \hat{\bm{\beta}}=\arg\max_{\bm{\beta}}\left\lbrace l(\bm{\beta})\right\rbrace.
\end{equation*}

In the context of high-dimensional survival analysis, where the number of covariates ($p$) far exceeds the sample size ($n$), traditional estimation approaches face critical challenges. This high-dimensional scenario creates an ill-posed mathematical problem for the standard Cox regression model where the design matrix becomes rank-deficient, resulting in infinite possible solutions for the regression coefficient vector $\bm{\beta}$. These theoretical limitations render unpenalized regression statistically infeasible in high-dimensional settings. Moreover, even when a solution can be computed, it typically results in severe overfitting.

To address these fundamental challenges, several techniques, commonly referred to as penalization or regularization methods, introduce constraints that ensure a unique solution. These include ridge (Hoerl $\&$ Kennard, \cite{Hoerl_1970}), lasso (Tibshirani, \cite{Tibshirani_1996}) and elastic net (Zou $\&$ Hastie, \cite{Zou1}) penalties among many others. Penalization methods are based on a bias-variance trade-off. By introducing a penalty term in the optimization, the variance of the estimated coefficients can be effectively reduced at the cost of an increased bias. This bias-variance trade-off can lead to more stable and interpretable models, particularly in high-dimensional settings where overfitting is a concern. These penalized models are based on minimizing
\[-l(\bm{\beta})+\sum_{i=1}^p P_{\lambda}(\beta_i),\]
where $l(\bm{\beta})$ is the log partial likelihood function given in Equation \eqref{loglik} and $P_{\lambda}(\cdot)$ is the penalty function with $\lambda>0$ being the regularization parameter that controls the penalty strength. 

The choice of the penalization method is heavily dependent on the specific characteristics of the problem at hand. In the context of the motivating dataset — a high-dimensional genetic setting with a very small number of observations - lasso and adaptive lasso methods are particularly appropriate choices. This research focuses specifically on these two penalization approaches due to their variable selection capabilities, which are essential in high-dimensional settings where identifying the most influential predictors among numerous covariates is paramount.

\subsection{Lasso}
The Least Absolute Shrinkage and Selection Operator (lasso), introduced by Tibshirani \cite{Tibshirani_1996} and later extended to survival analysis (Tibshirani, \cite{Tibshirani_1997}), simultaneously performs coefficient estimation and variable selection. In the context of Cox regression models, the lasso estimator is defined as
\begin{equation*}
\hat{\bm{\beta}}_{lasso}=\arg\min_{\bm{\beta}}\left\lbrace-l(\bm{\beta})+\lambda\sum_{i=1}^p|\beta_i|\right\rbrace,
\end{equation*}
where \(\lambda\) is the the regularization parameter controlling the penalty strength and \(\sum_{i=1}^p|\ \beta_i\ |\) represents the \(L_1\)-norm penalty. The key property of lasso lies in its ability to shrink some coefficients exactly to zero, effectively removing the corresponding variables from the model. This sparsity-inducing property makes lasso particularly valuable in high-dimensional analysis, where interpretability is fundamental.

However, despite its advantages and widespread adoption, lasso exhibits noteworthy limitations. Fan and Li \cite{Fan} demonstrated that lasso introduces excessive bias in the estimation of large coefficients. Specifically, in variable selection scenarios, lasso can be inconsistent, potentially including irrelevant variables while excluding truly significant ones and assigning coefficient values far from the true values (Zou \cite{Zou}).

\subsection{Adaptive lasso}
To address these fundamental limitations, in a seminal work Zou \cite{Zou} proposed the adaptive lasso for linear regression , which was subsequently extended to survival analysis by Zhang and Lu \cite{Zhang}. The adaptive lasso estimator for Cox regression is defined as
\begin{equation*}
\hat{\bm{\beta}}_{a.lasso}=\arg\min_{\bm{\beta}}\left\lbrace-l(\bm{\beta})+\lambda\sum_{i=1}^p w_{i}|\beta_i|\right\rbrace,
\end{equation*}
where \(w=\left(w_1,\ldots,w_p\right)^T\) is a vector of positive weights associated to each regression coefficient. These weights are typically derived from an initial estimator, with larger weights assigned to coefficients likely to be zero and smaller weights to significant coefficients. This adaptive penalization scheme effectively addresses the bias issue inherent in standard lasso.

\begin{figure}[ht]
    \includegraphics[width=0.5\linewidth]{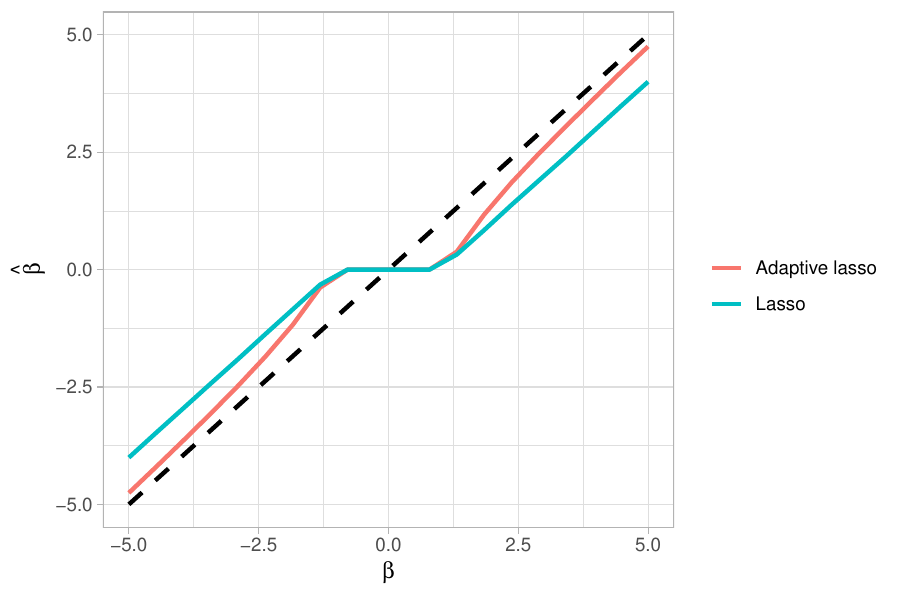}
    \centering
    \caption{Thresholding functions for lasso and adaptive lasso penalties, illustrating how each method shrinks coefficient estimates relative to their true values.}\label{fig:comparison}
\end{figure}

Figure \ref{fig:comparison} illustrates the distinct behavior of lasso (blue) and adaptive lasso (red) thresholding functions. Both approaches effectively shrink small coefficients toward zero, but their treatment of larger coefficients differs substantially. While lasso maintains a constant bias for large coefficients, adaptive lasso progressively reduces this bias as coefficient magnitude increases, resulting in estimates closer to the true values for important predictors.

The main theoretical advantage of adaptive lasso over lasso is its satisfaction of the oracle property under appropriate conditions. Following Fan and Li \cite{Fan}, an estimator possesses the oracle property if it is able to asymptotically identify the correct subset of non-zero coefficients as the sample size increases and achieves the same asymptotic distribution as if the true model were known in advance. Zou \cite{Zou} proved that adaptive lasso satisfies the oracle property when weights are derived from $\sqrt{n}$-consistent initial estimators in fixed-dimensional settings. However, this theoretical guarantee has not been extended to high-dimensional scenarios where $p \gg n$.

The weight specification is crucial for adaptive lasso's performance. Zhang $\&$ Lu \cite{Zhang} proposed setting \(w_i=1/|\hat{\beta_i}|\) for \(i=1,\ldots,p\) with \(\hat{\beta_i}\) being the estimator of \(\beta_i\) obtained by maximizing the unpenalized log partial likelihood function \(l(\bm{\beta})\). While effective in low-dimensional settings, this approach becomes problematic in high-dimensional settings where reliable estimation of unpenalized coefficients is infeasible.

A key contribution of this work is the development and evaluation of novel weight calculation strategies specifically designed for high-dimensional survival analysis. Building upon the foundation established by Méndez-Civieta \textit{et al}. \cite{Mendez}, we explore alternative weight calculation methods that leverage machine learning and statistical methodologies to overcome the limitations of conventional approaches in high-dimensional settings. While the theoretical guarantees of the oracle property do not extend to high-dimensional scenarios where $p \gg n$, our proposed weight calculation alternatives demonstrate empirically strong performance in variable selection and prediction accuracy, offering practical solutions for researchers working with extremely limited sample sizes relative to the number of predictors.

\section{Proposed weight calculation methods}\label{weights}

The specification of weights for adaptive lasso is key for both accurate variable selection and coefficient estimation. In low-dimensional settings where $n > p$, approaches such as those proposed by Zhang \& Lu \cite{Zhang} or Zou \cite{Zou} utilize unpenalized coefficient estimates to derive these weights. However, these methods become infeasible in high-dimensional scenarios where $p \gg n$. This  limitation requires alternative weight calculation strategies specifically designed for high-dimensional survival analysis. 

We propose and evaluate four distinct methodologies that overcome these challenges while preserving the desirable properties of adaptive lasso. These approaches leverage different statistical and machine learning techniques to approximate unpenalized coefficient values or identify important variables through alternative means. The parameter $\gamma$ introduced in the subsequent equations serves as a secondary tuning parameter that controls the adaptive weighting strength, where $\gamma = 0$ corresponds to lasso (equal weights), while $\gamma = 1$ represents the conventional adaptive lasso formulation. While $\gamma = 1$ is typically used in practice, we consider $\gamma \in [0.2, 2]$ to allow flexibility in the degree of adaptivity.

\subsection{Ridge weights}

Ridge regression, introduced by Hoerl \& Kennard \cite{Hoerl_1970}, offers a  suitable initial estimator for high-dimensional settings because it yields unique solutions even when $p \gg n$. Ridge regression is a penalization method that applies an $L_2$-penalty to the coefficients, effectively handling multicollinearity and ensuring computational stability. When applied to Cox regression, the ridge estimator is defined as:
\begin{equation*}
\hat{\bm{\beta}}^{(ridge)}=\arg\min_{\bm{\beta}}\left\lbrace-l(\bm{\beta})+\lambda\sum_{i=1}^p\beta_i^2\right\rbrace.
\end{equation*}

Ridge regression is especially appropriate for weight calculation because it is a root-$n$ consistent estimator that would satisfy the oracle property in low-dimensional settings, providing theoretical justification for its use as a basis for adaptive weights. While this consistency cannot be guaranteed in high-dimensional scenarios, ridge regression still provides stable coefficient estimates that reflect the relative importance of predictors.

Given the ridge estimates $\hat\beta^{(ridge)}_i$ for $i=1,\ldots,p$, the weights for adaptive lasso are computed as
\begin{equation}\label{Ridge}
w_i^{ridge}=\frac{1}{|\hat\beta^{(ridge)}_i|^{\gamma}}, \quad i=1,\ldots,p,
\end{equation}
where $\gamma$ controls the degree of penalization differentiation. Larger $\gamma$ values result in more aggressive differentiation between coefficients of important and unimportant variables, with small coefficients receiving substantially larger penalties than large coefficients.

\subsection{PCA weights}

Principal Component Analysis (PCA) offers an alternative approach to handling high dimensionality by reducing the data to a lower-dimensional representation that preserves the majority of the original variation. Unlike ridge regression, which shrinks coefficients but retains all variables, PCA creates a new coordinate system of uncorrelated components through linear combinations of the original predictors (Jolliffe \& Cadima, \cite{Jollife}).

This dimension reduction approach is particularly valuable in high-dimensional survival analysis because it allows us to circumvent the ill-posed problem by fitting a Cox model on a manageable number of components rather than the original high-dimensional feature space. The components capture the most important patterns in the data while discarding redundant or noisy dimensions.

Following Mendez-Civieta \textit{et al.} \cite{Mendez}, the PCA weight calculation procedure decomposes the covariate matrix $X$ of dimension $n \times p$ into $X=SP^T$, with $S$ being the ($n \times p$) matrix of scores and $P$ being the ($p \times p$) matrix of loadings. The procedure then (1) Selects the $r<\text{rank}(X)$ components that explain approximately $95\%$ of the total variability. (2) Fits a Cox regression model using the reduced score matrix $S_r$ as covariate matrix, which yields the estimated coefficients $\hat{\bm{\beta}}_r$. (3) Maps these coefficients back to the original variable space through $\hat{\bm{\beta}}^{(PCA)}=P_r\hat{\bm{\beta}}_r$ where $P_r$ is the matrix of loadings for the $r$ selected components. The weights for each variable are then calculated as:
\begin{equation}\label{PCA}
w_i^{PCA}=\frac{1}{|\hat\beta^{(PCA)}_i|^{\gamma}}, \quad i=1,\ldots,p,
\end{equation}
where $\gamma$ serves the same function as in the ridge approach.

This method effectively leverages the dimension-reducing capabilities of PCA to obtain meaningful coefficient estimates in high-dimensional settings, capturing the most important underlying patterns in the data while reducing noise.

\subsection{Univariate Cox regression weights}

The univariate approach offers a computationally efficient and conceptually straightforward alternative to multivariate modeling when calculating adaptive weights. Inspired by Belhechmi \textit{et al.} \cite{Bel}, this method sidesteps the challenges of multivariate high-dimensional estimation by examining each predictor's individual relationship with survival.

The motivation behind this approach is both practical and theoretical. From a practical point of view, univariate models are always estimable. From a theoretical perspective, while univariate models cannot capture complex variable interactions, they can effectively identify marginally important predictors, constituting a reasonable starting point for variable selection.

This procedure fits separate univariate Cox models for each covariate $X_j$, obtaining coefficient estimates $\hat{\beta}_j^{(Uni)}$ for $j \in \{1,\ldots,p\}$. The adaptive lasso weights are then computed as:
\begin{equation}\label{Uni}
w_i^{Uni}=\frac{1}{|\hat\beta^{(Uni)}_i|^{\gamma}}, \quad i=1,\ldots,p.
\end{equation}

This approach provides a computationally feasible method to approximate the full multivariate relationships while ensuring numerical stability in high dimensional settings.

\subsection{Random Survival Forest weights}

Modern machine learning methods capture complex, non-linear relationships in high-dimensional data, but often lack the variable interpretability, crucial for many clinical and scientific applications. To bridge this gap, we propose a novel weight calculation approach based on a machine learning methodology while maintaining the interpretability of penalized regression models.

Random Survival Forest (RSF), introduced by Ishwaran \textit{et al.} \cite{Ish}, is a non-parametric ensemble method that extends Breiman's random forest algorithm \cite{brieman1} to right-censored survival data. Survival trees are built by partitioning the data to form groups of subjects who are similar according to the survival outcome given by the Nelson-Aalen's cumulative hazard function (see Aallen, \cite{Aallen}; Nelson, \cite{Nel}). The ensemble is then a cumulative hazard function formed by averaging individual trees. See Lee $\&$ Lim, \cite{Seu}; Pickett \textit{et al.}, \cite{Pic} for applications of this methodology. By constructing an ensemble of survival trees, RSF naturally handles high-dimensional data, captures non-linear relationships and interactions, and provides variable importance measures that quantify each predictor's contribution to the model's performance. The most common variable importance measure is the Breiman-Cutler variable importance (VIMP) (Breiman, \cite{Brieman}), also known as permutation importance. For each variable $i$, VIMP quantifies the decrease in prediction accuracy when the variable's effect is nullified through permutation, with higher values indicating greater importance.

The RSF-based weight calculation leverages these variable importance measures to inform the adaptive lasso weights. Specifically, after fitting a Random Survival Forest, we extract VIMP. The weights for adaptive lasso are then calculated as
\begin{equation}\label{RSF}
w_i^{RSF}=\frac{1}{|\hat I_i|^{\gamma}}, \quad i=1,\ldots,p,
\end{equation}
where $\hat I_i$ represents the variable importance measure for the $i$-th covariate.

This innovative approach combines the advantages of both paradigms: utilizing the predictive capabilities of machine learning algorithms to identify important variables, while maintaining the interpretability of penalized regression models. By incorporating non-linear relationships and interactions into the weight calculation process, RSF weights can potentially identify important predictors that might be missed by the linear methods underlying the other approaches.

\section{Model evaluation and selection}\label{model_eval}

In high-dimensional survival analysis, robust evaluation metrics and reliable parameter selection techniques are critical for statistical inference and clinical interpretation. This section presents a comprehensive framework addressing three interconnected aspects of model evaluation and selection in Cox regression. First, we discuss the selection of the penalty parameter $\lambda$ using cross-validation methods specifically adapted for survival data where standard approaches may be problematic due to censoring. Second, we introduce appropriate performance measures for Cox regression models, focusing on concordance-based metrics that accommodate right-censored data while avoiding dependence on the censoring distribution. Finally, we propose a novel approach for model selection that addresses the inherent instability of variable selection in high-dimensional settings, extending the methodology of Laria \textit{et al.} \cite{Laria} from classification problems to survival analysis. This integrated approach allows us to identify models that not only achieve high predictive accuracy but also provide consistent and interpretable variable selection across different data partitions—a critical consideration in genomic applications where biological interpretation depends on the stability of selected predictors.

\subsection{Penalty parameter selection}\label{sec:cv}

The selection of the penalty parameter $\lambda$ is essential for obtaining correct estimation and selection of the coefficients. K-fold cross-validation is the most common method for selecting $\lambda$ in penalized Cox regression (Dai $\&$ Breheny, \cite{Dai}). In K-fold cross-validation, the data $D$ is partitioned into $K$-folds $D_1,\ldots,D_K$. For each $k\in\{1,\ldots,K\}$, $L^{-k}$ denotes the partial likelihood constructed using the data $D-D_k$ and $L^k$ represents the partial likelihood constructed using $D_k$. The estimates obtained with $L^{-k}$ are denoted by $\hat\beta^{-k}.$ 

The basic cross-validated partial likelihood approach computes the cross-validation error (CVE) as follows: 
$$CVE=-2\sum_{k=1}^K l^k (\hat\beta^{-k}),$$
with $l^k(\hat\beta^{-k})$ being the log-partial likelihood calculated using $D_k$ evaluated at $\hat\beta^{-k}.$ While this approach is commonly used for logistic and linear regression, it can be problematic for Cox regression because there might be insufficient events in a fold, which are necessary to calculate the partial likelihood denominator in $l^k$.

Verweij \& Van Houwelingen \cite{Ver} proposed an alternative approach to address this problem that constructs the CVE using $l(\hat\beta^{-k})$ and $l^{-k}(\hat\beta^{-k})$:
\begin{equation*}\label{eqcv}
CVE=-2\sum_{k=1}^K \left(l(\hat\beta^{-k})-l^{-k}(\hat\beta^{-k})\right).
\end{equation*}

This method avoids the problem of insufficient observations in each fold, as $(l^k)$ is not directly involved in the error calculation. In our study, we adopt this alternative cross-validation approach for simulations in Section \ref{simul} and the case study in Section \ref{cstudy}. 

\subsection{Model performance measures}\label{cind}

Evaluating the performance of Cox regression models also presents challenges, as these models do not directly predict survival times but rather estimate hazard functions for each sample. Consequently, traditional evaluation measures such as mean squared error are not applicable. Instead, concordance-based metrics that assess how well the model ranks patients by risk are more appropriate for survival data.

The concordance index (C-index) introduced by Harrell \textit{et al.} \cite{Harrell} provides the foundation for survival model evaluation. The C-index quantifies the proportion of concordant pairs —those where  the individual with the higher predicted risk experiences the event earlier. For a pair of individuals $(i,j)$ with survival times $T_i<T_j$ concordance occurs when the risk score $\eta_i>\eta_j$, where $\eta_i = \bm{\beta}^T \bm{X}_i$ represents the linear predictor for individual $i$. Formally, the C-index is defined as
\begin{equation*}\label{c}
C=P(\eta_i>\eta_j|T_i<T_j, \delta_i=1)=P(\bm{\beta}^T \bm{X}_i>\bm{\beta}^T \bm{X}_j|T_i<T_j, \delta_i=1),
\end{equation*}
and is typically estimated as 
\begin{equation*}
    \hat C=\frac{\sum_{i\neq j}\delta_i\mathbb{I}(T_i<T_j)\mathbb{I}(\hat{\bm{\beta}}^T \bm{X}_i>\hat{\bm{\beta}}^T \bm{X}_j)}{\sum_{i\neq j}\delta_i\mathbb{I}(T_i<T_j)}
\end{equation*}\,
where $\hat{\bm{\beta}}$ is the vector of estimated coefficients from a Cox regression model, $\delta_i$ is the censoring indicator, $\bm{X}_k$ is the covariate vector for the $k$-th individual and $\mathbb{I}(\cdot)$ is the indicator function. The censoring indicator $\delta_i$ identifies usable pairs, which must satisfy one of two conditions: $(1)$ both $T_i$ and $T_j$ are uncensored, or $(2)$ $T_i<T_j$, $T_i$ is uncensored and $T_j$ is censored.

Several time-dependent estimators have been developed to address limitations in the standard C-index. Heagerty $\&$ Zheng \cite{Hea} proposed a truncated C-index that evaluates predictive accuracy only up to a specified time point $\tau$, which is particularly valuable when the tail of the survival function may be unstable due to censoring, and Pencina $\&$ D'Agostino \cite{Pencina} presented an estimator for this truncated index. The estimator provides a time-specific assessment but remains dependent on the censoring distribution. To overcome this dependency, Uno \textit{et al.} \cite{Uno} introduced an improved estimator that incorporates inverse probability censoring weights based on the Kaplan-Meier estimator of the censoring distribution. This approach yields a consistent, non-parametric estimator robust to different censoring patterns. Despite these improvements, both estimators require selecting an appropriate time threshold $\tau$, which can be subjective and does not provide a global assessment of model performance across the entire follow-up period.

For high-dimensional survival analysis with heavy censoring, the concordance probability estimate (CPE) proposed by G\"onen $\&$ Heller \cite{Gonen}, also known as the K-index, offers significant advantages. This measure depends solely on the regression parameters and covariate distribution, without requiring observed or censoring times. As stated by Heller $\&$ Mo \cite{Heller_2016}, for continuous, independent and identically distributed random variables ${(T_i, X_i)}$, an application of Bayes' theorem shows that the concordance probability equals $K(\beta)=P(T_i<T_j|\bm{\beta}^T \bm{X}_i>\bm{\beta}^T \bm{X}_j)$, which can be theoretically expressed as:
\begin{equation*}
    P(T_i>T_j|\beta^T X_j>\beta^T X_i) =\frac{{\int\int}\mathbb{I}(\beta^T X_j>\beta^T X_i)(1+\exp(\beta^T(X_i-X_j)))^{-1}dF(\beta^T X_i) dF(\beta^T X_j)}{{\int\int}\mathbb{I}(\beta^T X_j>\beta^T X_i)dF(\beta^T X_i) dF(\beta^T X_j)},
\end{equation*}
with $F$ being the distribution function of $\eta_1=\beta^TX_1$. This leads to the CPE estimator:
\begin{equation}\label{cpe}
    K_n(\hat{\beta})=\frac{2}{n(n-1)}{\sum\sum}_{i<j}\left\lbrace\frac{\mathbb{I}(\hat\eta_j<\hat\eta_i)}{1+e^{(\hat\eta_j-\hat\eta_i)}}+\frac{\mathbb{I}(\hat\eta_i<\hat\eta_j)}{1+e^{(\hat\eta_i-\hat\eta_j)}}\right\rbrace.
\end{equation}

The estimator proposed in Equation \eqref{cpe} is a function of the regression parameters and the covariate distribution and does not use the observed and censoring times. For this reason it is asymptotically unbiased and, under standard conditions, $n^{1/2}\left(K_n(\hat{\bm{\beta}}) - K(\bm{\beta})\right)$ is asymptotically normal with mean zero, providing a statistically sound basis for model comparison. Henceforth, the subscript $n$ from $K_n(\hat{\bm{\beta}})$ will be omitted when it is not necessary to highlight the sample size. Throughout the remainder of this paper, we use $K(\hat{\bm{\beta}})$ to denote the CPE metric for evaluating model performance.

\subsection{Best model selection}\label{best_model}

A critical challenge in high-dimensional survival analysis is the inherent instability of variable selection across different data partitions. The estimated coefficient vector $\hat{\bm{\beta}}$ often varies substantially depending on how the data is split into training and test sets —a problem particularly pronounced in settings where $p \gg n$. This instability stems from the fundamental nature of high-dimensional problems, where solutions are more data-dependent than in low-dimensional settings. Different partitions of the same dataset can yield different selected variables, creating significant challenges for the interpretability of results. 

This variability is especially problematic in genomic applications, where consistently identifying significant genes is crucial for meaningful biological interpretation of the results. When selected genes change with each data partition, downstream analysis becomes more complex, affecting the translation of statistical findings into clinical insights. 

To address this challenge, we propose a novel model selection procedure that builds upon and extends the methodology introduced by Laria \textit{et al.}, \cite{Laria} for classification problems, adapting it to the unique characteristics of survival analysis. This approach represents a key contribution of our work, offering a robust framework for variable selection in high-dimensional survival settings that significantly enhances stability and reproducibility.

The procedure begins by fitting N Cox regression models on different data partitions. For each iteration $k$ for $k=1,\ldots,N$, the data is randomly split into $\text{Train}_k$ and $\text{Test}_k$. A Cox regression model with the appropriate penalization is then fitted on $\text{Train}_k$, with the penalty parameter $\lambda$ determined through the cross-validation process described in Section \ref{sec:cv}. The model's performance is then evaluated on $\text{Test}_k$ using the K-index performance measure $K(\hat{\bm{\beta}}^{(k)})$ introduced in Section \ref{cind}. 

Once the N models are fitted and their performance is evaluated, we calculate an importance index $I_j$ for each covariate $X^j$ for $j=1,\ldots,p$:
\begin{equation}\label{imp}
I_{j}=\frac{\sum_{k=1}^N{\left|\hat\beta_{j}^{(k)}\right|K(\hat\beta^{(k)})}}{\max_j\left\lbrace\sum_{k=1}^N\left|\hat\beta_j^{(k)}\right|K(\hat\beta^{(k)})\right\rbrace},
\end{equation}
with $\hat{\bm{\beta}}^{(k)}$ being the coefficients estimated by the model at iteration $k$ and $K(\hat\beta^{(k)})$ being the associated K-index metric. The index from Equation \eqref{imp} weights each coefficient by the performance of its corresponding model, assigning the value 1 to the most important covariate.

Since the true number of significant variables is typically unknown in real-world scenarios, following the suggestion by Laria \textit{et al.}, \cite{Laria},  we focus on identifying the top $K=\lceil \sqrt{n/2}\rceil$ most important variables, where $n$ is the number of observations in the data. Given the importance index of the $K$ most important variables, we define a power index $p_k$ for each model $k$ as:
\begin{equation*}
    p_k=\frac{1}{\sum_{i=1}^K I_{(i)}}\frac{\sum_{j:I_{j}\leq I_{(K)}}I_j\left|\hat\beta_j^{(k)}\right|}{\sum_{j=1}^p\left|\hat\beta_j^{(k)}\right|},\text{ for $k=1,\ldots,N,$}
\end{equation*}
where $I_{(i)}$ represents the $i$-th largest importance index associated to the $i$-th most important covariate. Finally, the best model is selected as the one that maximizes the sum of the K-index and the power index:
\begin{equation*}
\hat{\bm{\beta}}=\hat{\bm{\beta}}^{(J)}, \text{ with }J=\mathop{\arg\max}\limits_{j=1,\ldots,N}\lbrace K(\hat{\bm{\beta}}^{(j)})+p_j\rbrace.
\end{equation*}

The variables selected in the best model, which are the ones with non-zero coefficients, are then used to fit the final model using the complete dataset.

This method for selecting the best model represents a carefully balanced approach that seeks to optimize both predictive accuracy and variable selection stability. By combining the K-index (which measures pure predictive performance) with the power index (which quantifies how well a model captures consistently important variables across partitions), we address the dual objectives of survival analysis: accurate prediction and reliable identification of key predictors. This approach is particularly valuable in genomic studies where both aspects are critical —researchers need models that not only predict patient outcomes accurately but also consistently identify the same set of biomarkers across different data samples. In this procedure, the importance index weights the variables and the power index weights the models, providing a composite score that favors models with strong predictive performance that simultaneously select variables identified as important across multiple data partitions. This novel contribution significantly extends the work of Laria \textit{et al.}, \cite{Laria} from classification problems to survival analysis, offering a robust solution to the challenge of variable selection instability in high-dimensional survival settings.

\section{Simulation study}\label{simul}

This section presents a comprehensive simulation study designed to evaluate the performance of lasso and adaptive lasso penalization methods for Cox regression models in high-dimensional survival data settings. We systematically compare these methods across a range of different simulation schemes that cover different censoring proportions, different covariate structures and sparsity patterns. Our aim is to determine whether the adaptive lasso, with the various weight calculation procedures proposed in Section \ref{weights}, outperforms lasso in terms of variable selection and prediction accuracy. By examining these methods under controlled conditions, we provide empirical evidence of the improvements offered by the methodologies proposed in sections \ref{weights} and \ref{model_eval} for high-dimensional survival analysis applications.

\subsection{Data generation process}\label{sec:dgp}

To evaluate the performance of the proposed penalized Cox regression methods, we require a Data Generation Process (DGP) that accurately reflects the relationship between covariates and survival times while permitting controlled levels of right-censoring. Unlike linear regression, where responses are directly related to covariates, Cox regression models the relationship through the hazard function, requiring an inversion method to generate survival times corresponding to a specified baseline hazard and covariate effects.

Survival times were simulated based on the relationship between hazard and survival distributions detailed by Bender \textit{et al.} \cite{Ben}. For a given baseline hazard function $h_0(t) > 0$ and its corresponding cumulative hazard $H_0(t) = \int_{0}^t h_0(s)ds$, the survival time $T$ for an individual with covariate vector $X$ and regression coefficients $\beta$ follows the same distribution as $H_0^{-1}(-\log(U)\exp(-\beta^T X))$, where $U$ is a standard uniform random variable, $U \sim U[0,1]$. Therefore, simulating survival times requires specifying (1) the baseline cumulative hazard function $H_0(t)$, (2) the true regression coefficients $\beta$, and (3) generating random samples for the covariates $X$ and the uniform variable $U$.

In this study, we adopted a Weibull distribution for the baseline hazard, characterized by a shape parameter $\alpha > 0$ and a scale parameter $\rho > 0$, such that $h_0(t) = \alpha \rho^{-\alpha} t^{\alpha-1}$. The corresponding cumulative hazard function is $H_0(t) = (t/\rho)^\alpha$. Applying the inversion method, survival times $T$ were sampled from the distribution:
\begin{equation*}
    T = \rho\left(\frac{-\log(U)}{\exp(\beta^T X)}\right)^{1/\alpha}.
\end{equation*}

Introducing right-censoring at predefined rates was achieved by adapting the methodology of Wan \cite{Wan}. Wan's original approach, however, was developed for independent covariates, a condition often violated in genomic datasets where predictor correlations are common. To realistically simulate such scenarios, which represent a frequent challenge in practice, we developed an adaptation of Wan's method capable of handling the correlated covariate structures considered in our simulations.

Assuming the Weibull baseline hazard as specified above, the conditional density function of the survival time $T$ given covariates $X_i$ for subject $i$ is:
{\small
\begin{equation*}
f(t|X_i)=\frac{\alpha}{\left(\rho\exp(-\beta^TX_i/\alpha)\right)^{\alpha}}t^{\alpha-1}\exp\left(\frac{-t}{(\rho\exp(-\beta^TX_i/\alpha)^{\alpha}}\right)=\frac{\alpha}{\lambda_i^{\alpha}}t^{\alpha-1}\exp\left(\left(\frac{-t}{\alpha}\right)^{\alpha}\right),
\end{equation*}}
where $\lambda_i=\exp(-\beta_0/\alpha-\beta^{T}/\alpha X_i)$ and $\beta_0=\log(\rho^{-\alpha})$. In high-dimensional settings, it is simpler to work with the distribution of the derived variable $\lambda_i$. Depending on the distribution of the covariates, different expressions for $f_{\lambda_i}(\cdot)$ are obtained. Here the covariates are generated from a multivariate normal distribution with mean vector $0$ and covariance matrix $\Sigma$, $X\sim N(0,\Sigma)$, and based on this, $\lambda_i$ follows a log-normal distribution given by:
\begin{equation*}
    \lambda_i\sim \log N\left(-\beta_0/\alpha (\beta^T/\alpha)\Sigma(\beta/\alpha)\right),
\end{equation*}
with density function:
\begin{equation*}
    f_{\lambda_i}(u)=\frac{1}{u\sqrt{2\pi}\sqrt{(\beta^T/\alpha)\Sigma(\beta/\alpha)}}\exp\left(-\frac{(ln(u)+\beta_0/\alpha)^2}{2(\beta^T/\alpha)\Sigma(\beta/\alpha)}\right).
\end{equation*}
Let $G$ be the random variable that describes the right-censoring time, we assume that $G$ follows a Weibull distribution sharing the same shape parameter $\alpha$ as the baseline hazard, but having a potentially different scale parameter $\nu$ calibrated to achieve a target censoring proportion $\theta$. The probability that the $i-$th individual is censored ($\delta_i=0$) is derived as:

\begin{equation*}
P(\delta=0|\lambda_i,\alpha,\nu) =P(G\leq T\leq +\infty , 0\leq G\leq +\infty|\lambda_i,\alpha,\nu )= \frac{1}{1+(\nu/\lambda_i)^{\alpha}}
\end{equation*}
Then, the value of $\nu$ conditional on achieving a specific censoring proportion $\theta$ is obtained by solving $\gamma(\nu| \theta)=0$, where $\gamma(\nu| \theta)$ is the difference between the censoring rate and the censoring proportion $\theta$ and is given by:
\begin{equation*}
    \begin{split}
        \gamma(\nu| \theta)&=P(\delta = 0| \nu)-\theta \\
        &=\int_{0}^{+\infty}P(\delta = 0| u,\nu) f_{\lambda_i}(u) du-\theta\\
        &=E\left(\frac{1}{1+(\nu/\lambda_i)^{\alpha}}\right)-\theta,
    \end{split}
\end{equation*}
This expectation is generally intractable analytically but can be reliably estimated by the Strong Law of Large Numbers. Once the value of $\nu$ is determined numerically, censoring times $G$ are simulated from the Weibull($\alpha$,$\nu$). The observed time for each subject is then $T^* = \min(T, G)$, and the event indicator is $\delta = \mathbb{I}(T \le G)$.

The Weibull parameters used in our simulations ($\alpha=1.0032$, $\rho=320.7223$) were derived empirically by fitting a Weibull model to the real TNBC survival data analyzed in Section \ref{cstudy}. The goodness-of-fit of this Weibull baseline was confirmed using the modified Kolmogorov-Smirnov test for right-censored data (Fleming \textit{et al.} \cite{Fleming}). This choice ensures that the simulated baseline hazard characteristics are relevant to the motivating application.

To validate the effectiveness of this adapted censoring mechanism, particularly its ability to handle correlated covariates, the value of the scale parameter $\hat\nu$ is estimated for target censoring proportions ($\theta = 0.2, 0.4, 0.6, 0.8$) in two distinct scenarios: one with $4000$ independent normally distributed covariates and another with $4000$ correlated normally distributed covariates (using the block structure defined later). For each scenario and target censoring proportion $\theta$, we generated 100 replicate datasets and calculated the empirical censoring proportion $\hat\theta$ in each dataset. Table \ref{cens} compares the desired censoring proportion $\theta$ with the 
mean  and standard deviation (in parenthesis) of the empirical censoring proportions across the 100 replications. The close distance between target and observed proportions, along with low standard deviations, confirms that this procedure accurately and stably achieves the desired censoring levels for both independent and correlated high-dimensional covariates.

\begin{table}[ht]
    \small
    \centering
    \begin{adjustbox}{max width=\linewidth}
    \begin{tabular}{c|c|c|c} 
    \hline
        Scenario & Target $\theta$ & Estimated $\hat\theta$ & Estimated $\hat\nu$ \\ \hline
         & 0.20 & 0.200 (0.028) & 2270.892 \\ 
         & 0.40 & 0.398 (0.033) & 576.027 \\ 
         & 0.60 & 0.599 (0.037) & 178.668 \\ 
        \multirow{-4}{*}{$p=4000$ independent covariates} & 0.80 & 0.800 (0.026) & 45.278 \\ \hline
         & 0.20 & 0.209 (0.027) & 2273.393 \\ 
         & 0.40 & 0.399 (0.040) & 575.834 \\
         & 0.60 & 0.596 (0.035) & 178.752 \\ 
        \multirow{-4}{*}{$p=4000$ correlated covariates} & 0.80 & 0.797 (0.025) & 45.257 \\ \hline
    \end{tabular}
    \end{adjustbox}
    \caption{Validation of the censoring mechanism: Estimated scale parameter $\hat{\nu}$ for the Weibull($\alpha,\nu$) censoring distribution and resulting empirical censoring proportion $\hat\theta$ (mean (standard deviation) over 100 replicates) for target proportion $\theta$ under independent and correlated covariate structures.}\label{cens}
\end{table}

\subsection{Simulation design and parameters}\label{subsec:simulation_design}

This simulation study systematically investigates the performance of penalized Cox regression methods across a wide range of scenarios. For the covariate generation, we simulate from a multivariate normal distribution $X\sim N(0,\Sigma)$, where the covariance matrix $\Sigma$ varies according to three dependence scenarios: (1) Independent covariates: $\Sigma=\mathbb{I}_{nxp}$, (2) autoregressive correlation: $\Sigma_{i,j}=0.5^{|i-j|}$, as proposed by Tibshirani \cite{Tibshirani_1997} and (3) block dependence: $\Sigma_{i,j}=0.5\cdot\mathbb{I}(mod_{10}(i)=mod_{10}(j))$, following Freijeiro \textit{et al.} \cite{Freijeiro}. 

For each of the three covariate dependence structures described previously, we examine the following parameters:
\begin{itemize}
    \item Number of covariates $p$: $p=150,600,4000.$
    \item Number of non-zero coefficients $\varphi$: $\varphi=10,30,100$
    \item Coefficient values: the options considered are all non-zero coefficients being equal to $0.5$ or all non-zero coefficients taking values ranging from $1$ to $10$.
    \item Right-censoring proportion $\theta$: the proportion of right-censoring data takes values $\theta = 0, 0.2, 0.4, 0.6, 0.8 $. 
    \item Power $\gamma$: the adaptive weights obtained from each of the weight calculation procedures from Section \ref{weights} are powered to a parameter $\gamma$ that takes in the interval $[0.2, 2]$.
\end{itemize}

For each parameter combination, we generate data with a sample size of $400$ observations, split equally into training and testing sets of $200$ observations each. This training sample size reflects the dimensions expected in the case study presented in Section \ref{cstudy}. To ensure the stability of the results, we generate $100$ independent datasets for each scenario and report averaged performance metrics across these replications.

The models evaluated include:
\begin{itemize}
    \item Lasso: Cox regression with lasso penalty.
    \item Ridge: Cox regression with adaptive lasso penalty and ridge weights, $w_j^{ridge}$ as described in Equation \eqref{Ridge}.
    \item PCA: Cox regression with adaptive lasso penalty and PCA weights, $w_j^{PCA}$ as described in Equation \eqref{PCA}.
    \item Uni: Cox regression with adaptive lasso penalty and univariate weights, $w_j^{Uni}$ as described in Equation \eqref{Uni}.
    \item RSF: Cox regression with adaptive lasso penalty and weights computed with the importance index given by Random Survival Forests, $w_j^{RSF}$ as described in Equation \eqref{RSF}.
\end{itemize}

For each fitted model, the predicted risks are calculated on the test set. Then, the following performance evaluation metrics are computed:
\begin{itemize}
    \item True positive rate (TPR): proportion of correctly identified non-zero coefficients.
    \item False positive rate (FPR): proportion of incorrectly identified zero coefficients.
    \item F1 score: $F1=\frac{TP}{TP+\frac{1}{2}(FP+FN)}$, where FN is the false negative error, the number of incorrectly identified non-zero coefficients.
    \item Predicted risk: for each element on the test set, $\exp\left(\hat\beta^T X\right)\left(\exp\left(\beta^T X\right)\right)^{-1}$
    \item Coefficient estimation error: $\|\beta-\hat\beta\|_2,$ 
\end{itemize}

\subsection{Computational aspects}\label{subsec:computational_aspects}

The simulation study was implemented using the {\tt R} statistical software environment. In addition, the following specialized {\tt R} libraries were used:
\begin{itemize}
    \item \texttt{GofCens}: For the goodness-of-fit tests on the Weibull distribution parameters.
    \item \texttt{glmnet}: For implementing lasso, adaptive lasso, and ridge penalized Cox regression models, and performing 10-fold cross-validation to select the optimal penalty parameter $\lambda$
    \item \texttt{survival}: For fitting standard Cox regression models in the univariate weight calculation procedure and for general survival analysis functions
    \item \texttt{stats}: For principal component analysis in the PCA weight calculation procedure
    \item \texttt{randomForestSRC}: For Random Survival Forest implementation to calculate variable importance measures for RSF weights
\end{itemize}
The following github repository contains all the code needed to implement the formulas and the simulations from this section: 

\url{https://github.com/Pilargonzalezbarquero/penalized_cox}.

\subsection{Simulation results}

We now present the findings from our extensive simulation study, designed to evaluate the performance of lasso and the proposed adaptive lasso approaches (using Ridge, PCA, Univariate, and RSF weights) under various high-dimensional survival settings. We focus the discussion on scenarios with $p=4000$ covariates and $\varphi=30$ non-zero coefficients, varying the censoring proportion ($\theta$) and covariate dependence structures, as we think these can summarize the major findings of the simulation scheme, but the complete results are shown as part of the supplementary  material. The tables display the mean and the standard deviation (in parenthesis) of the TPR, FPR, F1 score, median ratio of predicted risk to true risk for predictive performance and the $L_2$-norm distance between estimated and true coefficients ($\|\beta-\hat\beta\|_2$). As a general guide, the best results will be associated to larger TPR, lower FPR, larger F1, median values close to $1$, and lower $\|\beta-\hat\beta\|_2$. The best results for each scenario are marked in \textbf{bold}.

A consistent observation across all scenarios is the decrease in performance for all methods as the proportion of censored data increases. This is expected, as higher censoring reduces the information available for model fitting. Notably, the adaptive lasso approach using Random Survival Forest (RSF) weights consistently underperformed compared to the other adaptive methods, particularly in variable selection accuracy (TPR and F1 score), as seen in the initial results for the independent covariate case (Table \ref{case1}, 0\% censoring). Consequently, the RSF-weighted adaptive lasso was excluded from further detailed comparisons in subsequent scenarios to maintain focus on the more promising approaches.

\begin{figure}[ht]
\includegraphics[width=\linewidth]{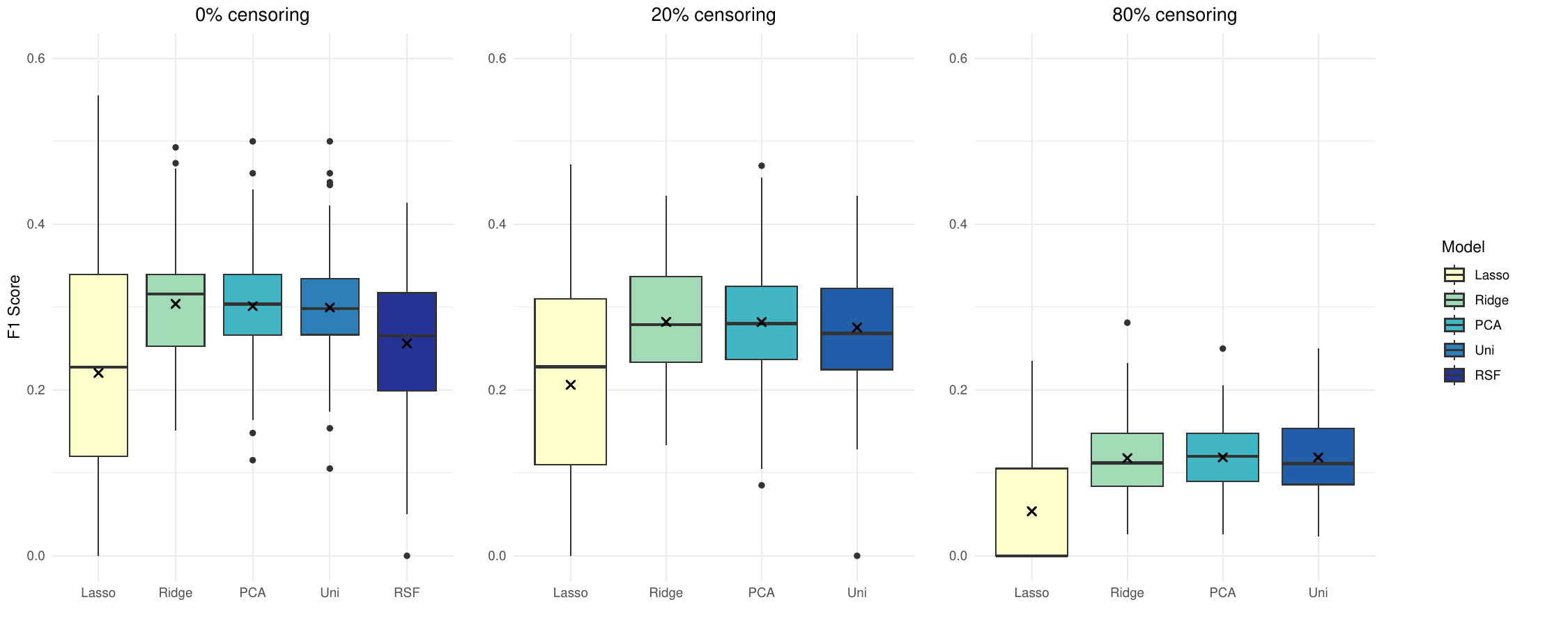}
\centering
\caption{F1 score results in case $1$ for different censoring proportions.}
\label{case1boxplot}
\end{figure}

In the scenario with independent covariates (Table \ref{case1}), the adaptive lasso methods using Ridge, PCA, and Univariate weights demonstrated superior variable selection performance compared to lasso. They achieved significantly higher TPR and F1 scores across different censoring levels, indicating a better ability to identify the true non-zero coefficients. While lasso showed a slightly lower FPR, its overall variable selection capability was much weaker. Interestingly, regarding prediction error (median predicted risk ratio) and coefficient estimation error ($\|\beta-\hat\beta\|_2$), all methods (Lasso, Ridge, PCA, Uni) performed similarly. This highlights a key advantage of the adaptive approaches: they significantly improve variable selection while maintaining, and sometimes improving, predictive and estimation accuracy compared to lasso. This advantage is clearly visualized in Figure \ref{case1boxplot}, which plots the F1 scores — a metric balancing precision and recall in variable selection — across different censoring levels. The figure highlights two critical findings regarding lasso's performance: it consistently yields the lowest median F1 scores, and perhaps more importantly, it exhibits substantially greater variability across simulation runs compared to the adaptive methods. Conversely, the adaptive approaches utilizing Ridge, PCA, and Univariate weights achieve significantly higher F1 scores while maintaining much lower variability, indicating more stable and reliable identification of the true non-zero coefficients.
\begin{table}[ht]
    \small
    \centering
    \begin{adjustbox}{max width=\linewidth}
    \begin{tabular}{c|c|c|c|c|c|c}
    \hline
        $\theta$ & Model ($\gamma$) & TPR & FPR & F1 score & Median & $\|\beta-\hat\beta\|_2$  \\ \hline
         & Lasso & 0.192 (0.156) & \textbf{0.002} (0.002) & 0.221 (0.149) & 0.978 & 2.708 (0.034) \tabularnewline
         & Ridge (1.2) & 0.417 (0.117) & 0.01 (0.003) & \textbf{0.304} (0.071) & 0.962 & 2.651 (0.063) \tabularnewline
         & PCA (1.2) & \textbf{0.419} (0.119) & 0.01 (0.003) & 0.301 (0.068) & 0.959 & \textbf{2.645} (0.064) \tabularnewline
         & Uni (1.2) & 0.418 (0.119) & 0.01 (0.003) & 0.299 (0.066) & 0.951 & 2.65 (0.064) \\ 
        \multirow{-5}{*}{0\%}& RSF (0.2) & 0.262(0.121) & 0.005 (0.002) & 0.256 (0.089) & \textbf{1.013} & 2.647 (0.052) \\ \hline
         & Lasso & 0.188 (0.146) & \textbf{0.003} (0.003) & 0.206 (0.133) & \textbf{0.988} & \textbf{2.707} (0.034) \\ 
         & Ridge (1) & \textbf{0.41} (0.112) & 0.011 (0.003) & \textbf{0.282} (0.069) & 0.924 & 2.665 (0.063) \\ 
         & PCA (1) & 0.398 (0.119) & 0.011 (0.003) & \textbf{0.282} (0.075) & 0.941 & 2.659 (0.061) \\ 
        \multirow{-4}{*}{20\%} & Uni (1) & 0.401 (0.119) & 0.011 (0.003) & 0.275 (0.073) & 0.947 & 2.666 (0.064) \\ \hline
         & Lasso & 0.044 (0.058) & \textbf{0.002} (0.003) & 0.054 (0.065) & 0.965 & \textbf{2.739} (0.024) \\ 
         & Ridge (1) & 0.157 (0.062) & 0.012 (0.003) & 0.118 (0.047) & 0.954 & 2.895 (0.117) \\ 
         & PCA (1) & \textbf{0.16} (0.062) & 0.012 (0.003) & \textbf{0.119} (0.043) & \textbf{0.979} & 2.885 (0.119) \\ 
         \multirow{-4}{*}{80\%}& Uni (1) & 0.158 (0.068) & 0.011 (0.002) & \textbf{0.119} (0.05) & 0.965 & 2.896 (0.109) \\ \hline
    \end{tabular}
    \end{adjustbox}
    \caption{Case 1: $4000$ \textbf{independent} covariates with $\varphi=30$ and non-zero coefficients equal to $0.5$.}\label{case1}
\end{table}

When introducing autoregressive correlation among covariates (Table \ref{case2}), the adaptive lasso methods (Ridge, PCA, Uni) maintained their superior performance over lasso. They consistently achieved higher TPR and F1 scores, indicating robust variable selection even with correlated predictors. Furthermore, in this correlated setting, the adaptive methods also tended to show slightly better results in terms of the median predicted risk ratio and coefficient estimation error compared to lasso, suggesting an improved handling of this type of dependency structure. The overall similarity in results between Table \ref{case1} and Table \ref{case2} suggests that the performance gains of adaptive lasso are robust to this form of correlation.

\begin{table}[ht]
    \small
    \centering
    \begin{adjustbox}{max width=\linewidth}
    \begin{tabular}{c|c|c|c|c|c|c}
    \hline
        $\theta$ & Model ($\gamma$) & TPR & FPR & F1 score & Median & $\|\beta-\hat\beta\|_2$ \\ \hline
         & Lasso & 0.184 (0.144) & \textbf{0.003} (0.002) & 0.209 (0.131) & 0.972 & 2.688 (0.049) \\ 
         & Ridge (1.2) & 0.387 (0.116) & 0.01 (0.003) & 0.283 (0.062) & 0.957 & \textbf{2.562} (0.1) \\ 
         & PCA (1.2) & \textbf{0.389} (0.109) & 0.01 (0.003) & \textbf{0.291} (0.07) & \textbf{1.005} & \textbf{2.562} (0.084) \\ 
        \multirow{-4}{*}{0\%} & Uni (1.2) & 0.387 (0.125) & 0.01 (0.003) & 0.28 (0.064) & 0.974 & 2.564 (0.098) \\ \hline
         & Lasso & 0.203 (0.143) & \textbf{0.003} (0.003) & 0.222 (0.125) & 0.979 & 2.681 (0.051) \\ 
         & Ridge (1) & \textbf{0.379} (0.112) & 0.011 (0.003) & \textbf{0.267} (0.064) & 0.992 & \textbf{2.572} (0.091) \\ 
         & PCA (1) & 0.371 (0.112) & 0.011 (0.003) & 0.266 (0.062) & 0.98 & \textbf{2.572} (0.087) \\ 
        \multirow{-4}{*}{20\%} & Uni (1) & 0.377 (0.113) & 0.011 (0.003) & 0.263 (0.06) & \textbf{0.995} & 2.573 (0.095) \\ \hline
         & Lasso & 0.054 (0.071) & \textbf{0.002} (0.003) & 0.063 (0.074) & 0.946 & \textbf{2.726} (0.027) \\ 
         & Ridge (1) & 0.161 (0.068) & 0.011 (0.002) & 0.122 (0.051) & 0.976 & 2.828 (0.1) \\ 
         & PCA (1) & \textbf{0.162} (0.061) & 0.011 (0.003) & \textbf{0.124} (0.045) & 0.949 & 2.809 (0.08) \\ 
        \multirow{-4}{*}{80\%} & Uni (1) & 0.159 (0.064) & 0.011 (0.003) & 0.123 (0.052) & \textbf{0.979} & 2.832 (0.104) \\ \hline
\end{tabular}
\end{adjustbox}
    \caption{Case 2: $4000$ \textbf{correlated} covariates with $\varphi=30$ and non-zero coefficients equal to $0.5$}\label{case2}
\end{table}

The scenario with block-correlated covariates (Table \ref{case3}) revealed differences among the adaptive methods. Here, the adaptive lasso using Ridge weights provided the best results across most metrics (TPR, F1 score, Median risk ratio, and $\|\beta-\hat\beta\|_2$), closely followed by the PCA-weighted approach. Both significantly outperformed lasso. However, the adaptive lasso with Univariate weights performed similarly to lasso in this setting, failing to show a clear advantage. This underscores that the choice of weighting strategy for adaptive lasso can have a large impact on the results, specially in complex high-dimensional settings. Here, Ridge and PCA weights appear more effective than Univariate weights when dealing with block-correlation structures. This highlights the importance of the contributions introduced in Section \ref{weights}.

\begin{table}[ht]
\small
\centering
\begin{adjustbox}{max width=\linewidth}
\begin{tabular}{c|c|c|c|c|c|c}
\hline
    $\theta$ & Model ($\gamma$) & TPR & FPR & F1 score & Median & $\|\beta-\hat\beta\|_2$ \\ \hline
     & Lasso & 0.345 (0.119) & 0.017 (0.003) & 0.191 (0.056) & 1.058 & 2.646 (0.066) \\ 
     & Ridge(1.4) & \textbf{0.476} (0.095) & \textbf{0.016} (0.002) & \textbf{0.261} (0.05) & \textbf{1.037} & \textbf{2.544} (0.106) \\ 
    & PCA(0.4) & 0.447 (0.107) & 0.018 (0.002) & 0.232 (0.05) & 1.06 & 2.575 (0.096) \\ 
     \multirow{-4}{*}{0\%}& Uni(0.2) & 0.34 (0.121) & 0.017 (0.003) & 0.187 (0.056) & 1.094 & 2.643 (0.072) \\ \hline
     & Lasso & 0.321 (0.102) & 0.018 (0.002) & 0.174 (0.05) & 1.061 & 2.653 (0.062) \\ 
     & Ridge (1) & \textbf{0.409} (0.088) & 0.017 (0.002) & \textbf{0.225} (0.048) & \textbf{1.044} & \textbf{2.606} (0.096) \\ 
     & PCA (1) & 0.363 (0.083) & \textbf{0.015} (0.002) & 0.215 (0.048) & 1.052 & 2.624 (0.09) \\ 
     \multirow{-4}{*}{20\%}& Uni (1) & 0.251 (0.095) & \textbf{0.015} (0.002) & 0.154 (0.054) & 1.053 & 2.688 (0.068) \\ \hline
     & Lasso & 0.111 (0.057) & 0.012 (0.002) & 0.08 (0.042) & 1.045 & \textbf{2.804} (0.063) \\ 
     & Ridge (1) & \textbf{0.125} (0.062) & 0.011 (0.001) & \textbf{0.094} (0.045) & \textbf{1.004} & 3.001 (0.136) \\ 
     & PCA (1) & 0.102 (0.055) & \textbf{0.01} (0.001) & 0.082 (0.044) & 1.038 & 2.985 (0.147) \\ 
    \multirow{-4}{*}{80\%} & Uni (1) & 0.088 (0.055) & \textbf{0.01} (0.002) & 0.072 (0.044) & 1.042 & 2.894 (0.1) \\ \hline
\end{tabular}
\end{adjustbox}
\caption{Case 3: $4000$ \textbf{correlated by blocks} covariates with $\varphi=30$ and non-zero coefficients equal to $0.5$}\label{case3}
\end{table}

Additional simulations in the supplementary material demonstrate similar findings for larger or smaller number of predictors ($p$), non-zero predictors ($\varphi$) and different non-zero predictor magnitudes. The clear pattern is that, whenever $p \gg n$, censoring severely impacts the identification of non-zero coefficients. However, adaptive lasso with Ridge or PCA weighting consistently reduces this effect more effectively than either the standard lasso or Uni weighting, showing more robust selection of important variables and more precise coefficient estimates for a broad range of simulation schemes. 

Further investigation of the block-correlated scenarios is shown in detail in the supplementary materials, in Tables S$53$-S$93$ and Figures S$1$-S$4$. These scenarios define a simulation scheme with $10$ groups with $3$ significant variables per group. A key finding is that all tested models exhibit consistent variable selection patterns across groups. As censoring percentage increases, the total number of selected covariates in each correlated block decreases across all models. While lasso and Uni occasionally fail to select all block groups in certain runs, Ridge and PCA approaches more reliably capture variables from every group of correlated predictors, even under heavy censoring. The percentage of correctly selected variables similarly decreases when the censoring percentage increases. Notably, under high censoring conditions (80\%), adaptive lasso with Ridge weights outperformed lasso in correctly selecting variables despite lasso selecting a larger total number of variables per group. 

Finally, we evaluated the model selection procedure proposed in Section~\ref{best_model} by applying it to a simulated dataset with $4000$ independent covariates, $\varphi=10$ non-zero coefficients and no censoring. In each method (lasso and adaptive lasso with Ridge, PCA, or Uni weights), we found that the importance index proposed in Equation \eqref{imp} successfully identified the 10 true non-zero covariates as the most important ones for all methods. Furthermore, the final "best" model selected by the procedure for each method correctly included these true important covariates. Notably, the coefficient estimates for these important variables in the adaptive lasso final models were consistently closer to their true values compared to the estimates obtained from lasso. This demonstrates that the proposed model selection procedure not only aids in identifying a stable and high-performing model but also enhances the accuracy of coefficient estimation for the selected important variables, particularly when used with the adaptive lasso methodologies. All these findings are included in the supplementary materials in Tables S121-S125.

\section{Survival analysis of breast cancer patients}\label{cstudy}

A detailed analysis of the genetic cancer dataset introduced in Section \ref{intro} is undertaken. This dataset, provided through a collaboration with the Gregorio Marañón General University Hospital (GMGUH), encompasses clinical and genetic information from patients from various hospitals in Spain and the National Institute of Neoplastic Diseases (INEN) in Lima, Peru. The study cohort consists of women diagnosed with triple-negative breast cancer (TNBC) who underwent neoadjuvant chemotherapy combined with docetaxel and carboplatin. 

TNBC represents $15-20\%$ of all breast cancer cases, being an aggressive subtype due to its high recurrence rates and rapid growth (Dent et al., \cite{cancer}). The aim of this study is to develop a statistical model capable of predicting patients survival times based on their available clinical and genetic information. Both the quality of the predictions and the selection of variables are essential for having a precise outcome.

The dataset is high-dimensional, containing $25$ clinical variables and $19571$ genetic variables. The genetic variables are interpreted as the counts of the messenger RNA expression for each observed gene and the clinical variables contain information such as tumor size, age at diagnosis, cancer staging or family history of cancer. For this study, we selected the $234$ patients for whom both clinical and genetic data are available, resulting in a high-dimensional scenario where $p \gg n$. The data exhibit $82\%$ right censoring, with $191$ censored observations and only $43$ observed events. The target variable for each patient is the number of months from the beginning of the chemotherapy treatment with docetaxel and carboplatin until death.

\subsection{Model fitting and results}

In this section, we aim to fit and compare different types of penalized Cox proportional hazard models with lasso and adaptive lasso penalties, evaluate their performance and compare their variable selection capabilities. For the adaptive lasso penalty, we apply the different types of weight computation methods defined in Section \ref{weights}, except for RSF weights, which were excluded based on their inferior performance in the simulation studies presented in Section \ref{simul}. Before fitting the models, the optimal penalty value $\lambda$ is determined through the cross-validation approach defined in Section \ref{sec:cv}.

An important challenge in high-dimensional survival analysis with small sample sizes is the instability of results across different train-test partitions of the data. To illustrate this issue, Table \ref{tab:my-table1} presents the results for the four Cox regression models (lasso and adaptive lasso with PCA, Ridge and Uni weights) evaluated on two different random train-test partitions of the data.The table shows that lasso selects only one variable in the first partition but 21 variables in the second partition. Similarly, the Uni weighting approach shows substantial variation, selecting 44 variables in Partition 1 versus 61 variables in Partition 2. The optimal model also changes between partitions: Ridge performs best in Partition 1 with $K(\hat{\beta})=0.883$, while Uni performs best in Partition 2 with $K(\hat{\beta})=0.901$.  This demonstrates how the selected variables and model performance can significantly vary depending on the specific data partition used, and clearly underscores the need for the novel approach to model selection presented in Section \ref{best_model}.

\begin{table}[ht]
\small
\centering
\begin{tabular}{c|c|c|c|c|c}
\hline
Partition & Model & min(CVE) & $K(\hat{\beta})$ & $\lambda$ & Num variables \\
\hline
\multirow{4}{*}{Partition 1} & Lasso & 10.918 & 0.616 & 0.127 & 1 \\
 & Ridge & 9.691 & 0.883 & 1.809 & 19 \\
 & PCA & 10.416 & 0.835 & 3.364 & 21 \\
 & Uni & 8.596 & 0.798 & 2.581 & 44 \\
\hline \hline
\multirow{4}{*}{Partition 2} & Lasso & 9.835 & 0.798 & 0.085 & 21 \\
 & Ridge & 8.859 & 0.894 & 1.056 & 41 \\
 & PCA & 9.061 & 0.783 & 12.558 & 13 \\
 & Uni & 8.033 & 0.901 & 1.144 & 61 \\
\hline
\end{tabular}
\caption{Comparison between Cox models with different penalty methods for two different partitions of the data real data.}
\label{tab:my-table1}
\end{table}

\begin{figure}[ht]
\includegraphics[height=0.2\textheight]{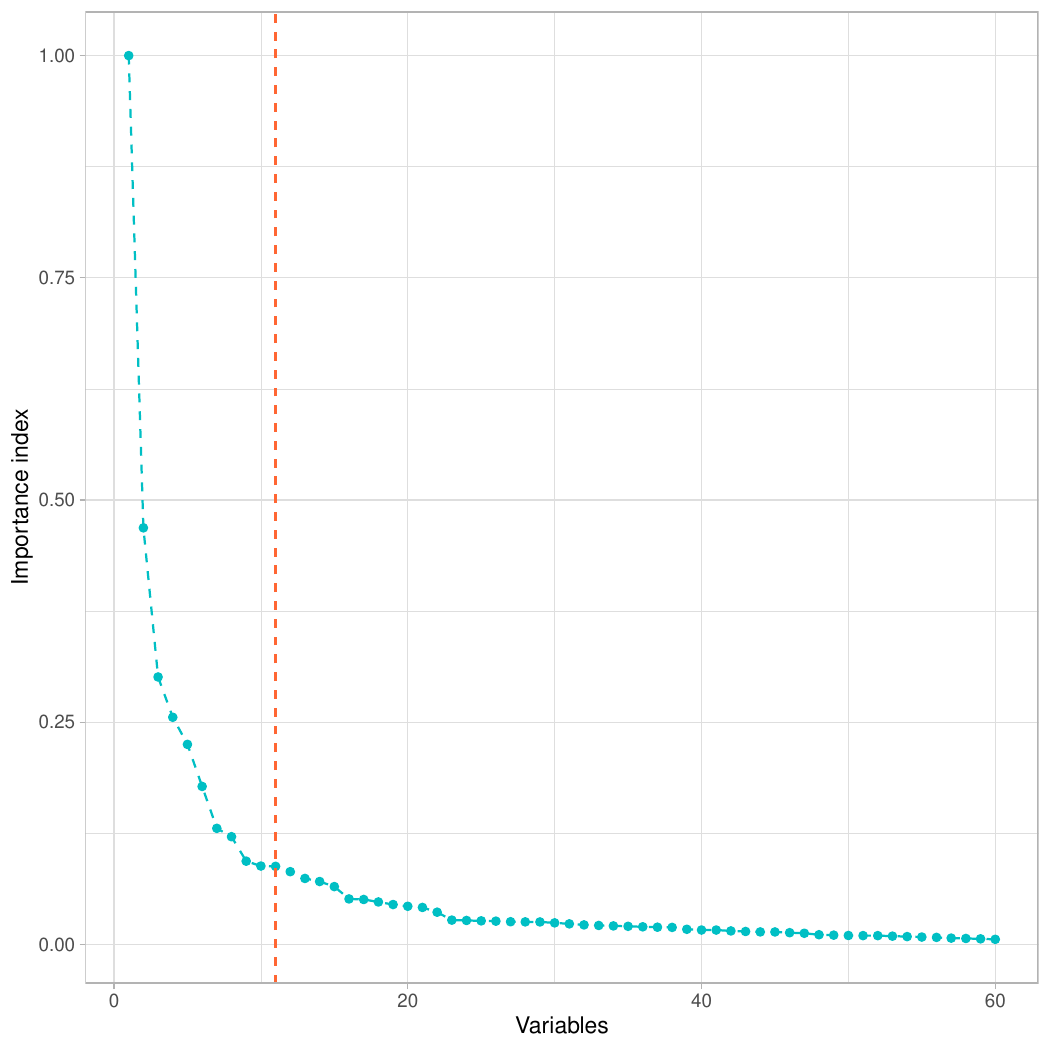}
\centering
\caption{Sorted variable importance indexes for adaptive lasso with PCA weights in the real data analysis.}
\label{fig:alasso}
\end{figure}

To address this variability, we implement the best model selection procedure proposed in Section \ref{best_model}. Section \ref{simul} and the supplementary materials show that Ridge and PCA weights provide the best results across a wide range of different simulation scenarios. For this reason, we focus our discussion primarily on Ridge and PCA results. The procedure involves creating 100 different random train-test partitions of the data with the train sets containing 200 observations and the test sets containing 34 observations. For each partition, we fit the penalized Cox regression models, compute their performance metrics, and record the selected variables. We then apply the importance index calculation detailed in Section \ref{best_model} to identify the most significant predictors across all partitions. Given our sample size of $n=234$, we calculate the power index for each model using the $K=\lceil \sqrt{n/2}\rceil=\lceil \sqrt{234/2}\rceil=11$ most important variables according to the importance index. Figure \ref{fig:alasso} displays the variables sorted by their importance index values for adaptive lasso with PCA weights, revealing a characteristic elbow pattern where importance values decline rapidly after the top few variables. This pattern, similar to the scree plots in other statistical methodologies, provides visual confirmation that a small subset of variables accounts for most of the predictive power in the model.

Following the methodology outlined in Section \ref{best_model}, we select the final model for each penalization approach by identifying the model that maximizes the sum of the K-index and the power index. Specifically, for each candidate model, we calculate its K-index (measuring predictive performance) and power index (measuring how well it captures the most important variables identified across all partitions). The model with the highest combined score is then fitted on the complete dataset using only the selected variables.

The best lasso model selects 10 variables out of which only one is clinical, {\tt n\_cat}, which measures tumor spread to lymph nodes. This variable was also  identified as the most important predictor according to the importance index. For the adaptive approaches, the best Ridge and PCA models select substantially more variables: 33 and 37 variables respectively. The PCA model selects primarily genetic variables, with only two clinical variables: {\tt n\_cat} and {\tt multicentric}. The Ridge model selects four clinical variables: {\tt n\_cat}, {\tt multicentric}, {\tt Hospital\_Universitario\_de\_Gran\_Canaria}, and {\tt Instituto\_Nacional\_Enfermedades \\ \_Neoplasicas}, where {\tt multicentric} indicates whether the patient has two or more tumors developing independently, and the last two variables identify the hospitals where patients received treatment.

Table \ref{tab:tab-3} analyzes the overlap between the sets of variables selected by each model. This analysis reveals that all three models share five genetic variables plus the clinical variable {\tt n\_cat}, suggesting these predictors are particularly robust indicators of survival in TNBC patients. Notably, the adaptive approaches (Ridge and PCA) demonstrate greater similarity to each other, sharing 9 common variables, compared to their overlap with the standard lasso approach. Lasso shares only 8 variables with Ridge (the 6 common to all plus 2 others) and 7 with PCA (the 6 common and one other). This pattern highlights one of the advantages of adaptive penalization approaches demonstrated in our simulation studies from Section \ref{simul}: they tend to provide more consistent variable selection.

\begin{table}[ht]
\small
\centering
\begin{tabular}{c|c}
\hline
Models & Common variables \\ \hline
Lasso $\&$ Ridge & \begin{tabular}[c]{@{}c@{}}n\_cat\\ ENSG00000137509\\ ENSG00000050030\\ ENSG00000176920\\ ENSG00000183336\\ ENSG00000139880\\ ENSG00000186352\\ ENSG00000167889\end{tabular} \\ \hline
Lasso $\&$ PCA & \begin{tabular}[c]{@{}c@{}}n\_cat \\ ENSG00000117620\\ ENSG00000137509\\ ENSG00000176920\\ ENSG00000183336\\ ENSG00000139880\\ ENSG00000167889\end{tabular} \\ \hline
\end{tabular}\qquad
\begin{tabular}{c|c}
\hline
Models & Common variables \\ \hline
Ridge $\&$ PCA & \begin{tabular}[c]{@{}c@{}}multicentric \\ n\_cat \\ ENSG00000101557\\ ENSG00000137509\\ ENSG00000099256\\ ENSG00000176920\\ ENSG00000183336\\ ENSG00000139880\\ ENSG00000167889\end{tabular} \\ \hline
Lasso $\&$ Ridge $\&$ PCA & \begin{tabular}[c]{@{}c@{}}n\_cat \\ ENSG00000137509\\ ENSG00000176920\\ ENSG00000183336\\ ENSG00000139880\\ ENSG00000167889\end{tabular} \\ \hline
\end{tabular}
\caption{Common variables in final models.}
\label{tab:tab-3}
\end{table}

In summary, our analysis demonstrates that the adaptive lasso approaches with Ridge and PCA weights provide more consistent variable selection than lasso in this high-dimensional survival analysis application. This consistency is evidenced by the higher number of common variables between Ridge and PCA models compared to their overlap with the lasso model. Additionally, the adaptive approaches identify more clinical variables, which may offer greater interpretability for medical researchers. These findings, combined with the superior predictive performance observed in our simulation studies, suggest that adaptive lasso with Ridge or PCA weights represents a promising approach for analyzing high-dimensional genetic survival data. The selected variables will be further investigated by the GMGUH research team to explore their biological mechanisms and potential clinical implications with TNBC and the chemotherapy treatment. 

\section{Discussion}\label{conclusion}

The primary objective of this research was to develop and evaluate high-dimensional pena\-lized Cox regression methods capable of handling the challenges posed by extremely large sets of predictors and high censoring rates. In particular, we proposed an extension of the adaptive penalization approach introduced by Mendez-Civieta \textit{et al.} \cite{Mendez} from quantile regression to survival contexts, and we incorporated a Random Survival Forest-based procedure as an alternative for adaptive weight calculation. We also introduced a new way to quantify the stability and relevance of selected predictors via an importance index and combined it with a power index to build a best model selection procedure that helps mitigate the variability of variable selection across multiple train-test partitions. Finally, we extended the Wan methodology \cite{Wan} to simulate survival data with predetermined right-censoring rates in correlated and high-dimensional scenarios.

The proposed methods were first analyzed through extensive simulations where adaptive lasso consistently demonstrated better variable selection accuracy than lasso, even under high censoring or strong correlations among predictors. Methods using Ridge or PCA for weights consistently showed the best variable selection and coefficient estimation. These results underscore the value of adaptive approaches in genomic applications, where a small number of truly relevant predictors may be hidden among thousands of potential covariates, and where data are often heavily censored.

In the genetic application to triple-negative breast cancer data  provided by the GMGUH the small sample size significantly impacted the performance of individual models. This issue was addressed using the best model selection method introduced in Section \ref{best_model}. Additionally, we observed that adaptive lasso models showed a more stable selection of variables, and selected more relevant clinical variables than lasso —including multicentricity, treatment center indicators, and extent of lymph-node involvement— alongside genetic components. By contrast, Lasso selected fewer common variables, including fewer clinical ones, and showed less stability across partitions. This difference can prove essential when interpretability is critical for downstream biological or clinical investigation.

In conclusion, adaptive lasso models, particularly with PCA and ridge weights, offer superior performance over lasso in high-dimensional survival analysis, both in synthetic and real data scenarios. Additionally, the proposed model selection procedure effectively mitigates variability issues due to different data partitions, enhancing the reliability of the selected models. 

Although these results showcase the advantages of adaptive lasso in a variety of high-dimensional survival contexts, there remain open questions to improve performance and interpretability. Future work could explore penalization methods that incorporate grouped structures such as sparse group lasso to better capture known relationships among genes, extending them to survival analysis. We also plan to develop alternative weight estimation procedures to improve the interpretation and performance of adaptive models. 

\section{Acknowledgements}
This research is part of the I+D+i projects PDC2022-133359 and PID2022-137243OB-I00 funded by  \\ MCIN/AEI/10.13039/501100011033 and European Union NextGenerationEU/PRTR. This initiative has also been partially carried out within the framework of the Recovery, Transformation and Resilience Plan funds, financed by the European Union (Next Generation) through the grant ANTICIPA and the ENIA 2022 Chairs for the creation of university-industry chairs in AI-AImpulsa: UC3M-Universia.

\bibliographystyle{plain} 
\bibliography{full_article.bib}      

\end{document}